\documentclass[iop]{emulateapj}       % for emulating 
\usepackage{multirow}
\usepackage{bm}
\usepackage{amssymb}
\usepackage{amsmath}
\usepackage{latexsym}
\usepackage{graphicx}
\usepackage{ifsym}
\usepackage{bm}
\usepackage{tabularx}
\usepackage{hyperref}
\usepackage{enumitem}

\newcommand{\about}{\mbox{$\sim$}}               % ~ (about) 
\newcommand{\Lsol}{\mbox{$L_\odot$}}             % Lsol
\newcommand{\hr}{\mbox{$^{\rm h}$}}                   % ^h
\newcommand{\mn}{\mbox{$^{\rm m}$}}                % ^m
\newcommand{\plus}{\mbox{$+$}}    % +
\newcommand{\kms}{\mbox{km s$^{-1}$}}                                    % km/s
\newcommand{\twelveCO}{\mbox{$^{12}$CO}}
\newcommand{\thirteenCO}{\mbox{$^{13}$CO}}
\newcommand{\CeighteenO}{\mbox{C$^{18}$O}}

\newcommand{\MHtwo}{M_{\mbox{\scriptsize{H}}_2}}
\newcommand{\MHone}{M_{\mbox{\scriptsize{HI}}}}
\newcommand{\Msol}{\mbox{$M_\odot$}}
\newcommand{\Mstar}{\mbox{$M_{*}$}}

\newcommand{\nd}{\nodata}

\shorttitle{GMCs in NGC 6946}
\shortauthors{Wu, Sakamoto, \& Pan}
\slugcomment{}

\begin{document}
\title{{\textbf {\large S\lowercase{ubmillimeter} A\lowercase{rray} \twelveCO~(2--1) I\lowercase{maging of the} NGC 6946 G\lowercase{iant} M\lowercase{olecular} C\lowercase{louds}}}}

\author{
Ya-Lin Wu$^1$,
Kazushi Sakamoto$^2$, and
Hsi-An Pan$^{2}$
}

\affil{$^1$Steward Observatory, University of Arizona, Tucson, AZ 85721, USA; yalinwu@email.arizona.edu\\
$^2$Academia Sinica, Institute of Astronomy and Astrophysics, Taiwan\\
{\it Published in ApJ}}

\begin{abstract}
\noindent
We present a \twelveCO~(2--1) mosaic map of the spiral galaxy NGC 6946 by combining data from the Submillimeter Array and the IRAM 30 m telescope. We identify 390 giant molecular clouds (GMCs) from the nucleus to 4.5 kpc in the disk. GMCs in the inner 1 kpc are generally more luminous and turbulent, some of which have luminosities $>$$10^6~\text{K km s}^{-1}~\text{pc}^{2}$ and velocity dispersions $>$10~\kms. Large-scale bar-driven dynamics likely regulate GMC properties in the nuclear region. Similar to the Milky Way and other disk galaxies, GMC mass function of NGC 6946 has a shallower slope (index $>-2$) in the inner region, and a steeper slope (index $<-2$) in the outer region. This difference in mass spectra may be indicative of different cloud formation pathways: gravitational instabilities might play a major role in the nuclear region, while cloud coalescence might be dominant in the outer disk. Finally, the NGC 6946 clouds are similar to those in M33 in terms of statistical properties, but they are generally less luminous and turbulent than the M51 clouds.\vspace{4pt}
\end{abstract}

\keywords{ 
        galaxies: individual (NGC 6946) --
        galaxies: ISM --        
        galaxies: spiral --
        ISM: clouds
       }

\section{Introduction}
\label{s.introduction}
Giant molecular clouds (GMCs) are the major gas reservoirs in disk galaxies, as well as the nursery of most stars. Their properties and life cycles likely regulate the initial mass function (e.g., \citealt{L81}). Our knowledge of star formation, the interstellar medium (ISM), and galaxy evolution ultimately hinges on a comprehensive understanding of GMCs.

Correlations between cloud size, luminosity, and velocity dispersion were first noticed by \cite{L79}. These correlations, or ``Larson laws,'' were established by \cite{S87}. Subsequent CO surveys have imaged GMCs from the Galactic center to the outer regions (e.g., \citealt{H01,H09,O01}). Over the past decade, advances in interferometry have enabled observers to probe GMCs in the Local Group and beyond (e.g., \citealt{RB05,R07,B08,W11,G12,R12,R15,DM13,C14,L15,P15b,U15}). Many studies have shown that extragalactic GMCs largely follow similar Galactic scaling relations, with some modifications possibly due to stellar feedback (e.g., \citealt{L15}), external pressure (e.g., \citealt{RB05}), or simply different viewing angles \citep{P16}.

Extragalactic GMC studies require data from interferometers, as well as single-dish telescopes to spatially and spectrally resolve a sufficient amount of clouds while conserving all the flux. In this paper, we study GMCs in the nearby gas-rich and moderately inclined galaxy NGC 6946 by combining our new \twelveCO~(2--1) observations from the Submillimeter Array\footnotemark[3] (SMA) and archive data from the IRAM 30 m dish \citep{L09}. This combination has great potential, because mapping a galaxy in \twelveCO~(2--1) is faster than other CO transitions for the same resolution and sensitivity \citep{S08}. Studies have found that molecular gas is more abundant than atomic gas in the inner 6 kpc \citep{CT07}, especially in the nuclear bar (e.g., \citealt{I90,RV95,Sakamoto99,S06}). Molecular gas can even be found outside the de Vaucouleurs radius $r_{25}$ \citep{B07}. Such a high molecular gas content is likely to drive the observed starburst \citep{TH83,E96}. In addition, NGC 6946 has been extensively investigated in different ISM phases (e.g., \citealt{H03,K03,K07,W08,H09}), facilitating a comparative study. Here we focus on the \twelveCO~(2--1) data and leave the comparison of cloud properties and star formation to a future paper. Table \ref{t.6946param} lists the properties of NGC 6946.

\footnotetext[3]{The Submillimeter Array is a joint project between the Smithsonian Astrophysical Observatory and the Academia Sinica Institute of Astronomy and Astrophysics, and is funded by the Smithsonian Institution and the Academia Sinica.} 

%%%%%%%%%%%%%%%%%
\begin{deluxetable}{@{}llc@{}}
%\tabletypesize{\scriptsize}
\tablewidth{\linewidth}
\tablecaption{NGC 6946 Parameters  \label{t.6946param} }
\tablehead{
	\colhead{\hspace{-27pt}Parameter}  &
	\colhead{\hspace{-32pt}Value}  &
	\colhead{Note}
}
\startdata  
R.A. (J2000)         & 20\hr34\mn52\fs355       & (1) \\ 
Decl. (J2000)       & \plus60\arcdeg09\arcmin14\farcs58 & (1) \\
Hubble type & SAB(rs)cd & (2) \\
$D$ [Mpc] & 5.5 & (3) \\
Scale. 1\arcsec~in pc & 26.7  \\
P.A.  [\arcdeg] & 243 & (4) \\
Incl. [\arcdeg] & 38 & (5) \\
$V_{\rm sys}$ [\kms] &  43 &  (5) \\
$r_{25}$ [kpc]  &  9.8   &  (6) \\
$M_B$ [mag]   &  $-20.61$ &  (6) \\
$L_{\rm 8-1000\mu m}$ [\Lsol] & $10^{10.2}$ & (7) \\
$\MHtwo$ [\Msol]   &   $4.0 \times 10^{9}$           &  (6)        \\
$\MHone$ [\Msol]   &   $6.3 \times 10^{9}$           &  (6)        \\
$\Mstar$ [\Msol]   &   $3.2 \times 10^{10}$          &  (6)        \\
SFR [\Msol~$\mbox{yr}^{-1}$]   &   3.24  &  (6)      
\enddata
\tablecomments{
(1) \cite{S06},
(2) \cite{RC3},
(3) \cite{T88},
(4) \cite{deBlok08},
(5) \cite{Boomsma08},
(6) \cite{L08},
(7) adjusted from the IRAS flux measurements in \cite{Sanders03} for the adopted distance of 5.5 Mpc.
}
\end{deluxetable}

\begin{deluxetable*}{@{}ccrclllccrcll@{}}
\tabletypesize{\scriptsize}
%\tablewidth{0pt}
\tablewidth{\linewidth}
\tablecaption{SMA Observations  \label{t.obslog} }
\tablehead{ 
         \colhead{No.} &
	\colhead{UT Date}  &
	\colhead{$N_{\rm p}$} &		
	\colhead{$N_{\rm ant}$} &	
	\multicolumn{2}{l}{Array Configuration} &
	\colhead{$L_{\rm baseline}$} &	
	\colhead{$\tau_{225}$} &
	\colhead{$\langle T_{\rm sys}\rangle$} &
	\colhead{$T_{\rm obs}$} &
	\colhead{Gain Cal.} &
	\colhead{Flux Cal.} &
	\colhead{Passband Cal.} 
	\\
	\colhead{ }  &	
	\colhead{ }  &
	\colhead{ } &
	\colhead{ } &	
	\colhead{name} &
	\colhead{pads} &
	\colhead{(m)} &	
	\colhead{} &
	\colhead{(K)} &
	\colhead{(hr)} &
	\colhead{ }  &	
	\colhead{ }  &
	\colhead{ } 
	\\
	\colhead{(1)}  &
	\colhead{(2)}  &
	\colhead{(3)} &
	\colhead{(4)} &
	\colhead{(5)} &
	\colhead{(6)} &
	\colhead{(7)} &
	\colhead{(8)} &
	\colhead{(9)} &
	\colhead{(10)} &	
	\colhead{(11)} &	
	\colhead{(12)} &
	\colhead{(13)} 
}
\startdata
1 & 2010 Jun 07 & 10 & 8 & CMP & 1,4,5,7,8,9,12,23        & 6--69  &  0.16 & 168 & 3.6 & J1849+670 & Uranus & 3C 454.3\\ 
&&&&&&&&&& J1927+739 & Titan & 3C 279 \\
2 & 2010 Sep 06 & 10 & 7 & EXT & 9,11,12,14,15,16,17    & 18--180  &  0.06 & 99 & 6.7 & J1849+670 & Uranus & 3C 454.3\\ 
&&&&&&&&&& J2038+513 & \nd & \nd \\ 
3 & 2010 Sep 07 & 10 & 7 & EXT & 9,11,12,14,15,16,17    & 21--176  & 0.06  & 101 & 5.1& J2015+371 & Uranus & 3C 454.3\\  
&&&&&&&&&& J2038+513 & Ganymede & \nd \\  
4 & 2010 Sep 08 & 10 & 7 & EXT & 9,11,12,14,15,16,17    & 22--175  & 0.05  &  97 & 4.8 & J2015+371 & Uranus & 3C 454.3\\   
&&&&&&&&&& J2038+513 & Callisto & 3C 84 \\
5 & 2010 Oct 27 & 10 & 7 & CMP & 1,4,5,7,8,9,12,23        & 7--69    & \about0.15 & 148 & 5.2 & J2015+371 & Uranus & 3C 454.3\\
&&&&&&&&&& J2038+513 & Callisto & 3C 84 \\
6 & 2012 Aug 05 &   2 & 6 & SUB & 1,2,3,4,5,6                   & 6--25    & \about0.25 & 182 & 5.2  & J1849+670 & Uranus & 3C 454.3\\
&&&&&&&&&& \nd & Neptune & \nd \\
7 & 2012 Aug 15 &   2 & 6 & SUB & 1,2,3,4,5,6                   & 6--25    & \about0.1   & 108 & 5.0  & J1849+670 & Uranus & 3C 454.3\\
&&&&&&&&&& \nd & Neptune & \nd 
\enddata
\tablecomments{
(3) Number of observed positions.
(4) Number of available antenna.
(5) SMA antenna configuration. CMP---compact, SUB---sub-compact, EXT---extended.
(6) Antenna locations. See \cite{Ho04} for a map with the numeric keys.
(7) Range of projected-length of baselines for NGC 6946.
(8) Zenith opacity at 225 GHz measured at the Caltech Submillimeter Observatory adjacent to the SMA.
(9) Median double sideband (DSB) system temperature toward NGC 6946.
(10) Total integration time on the galaxy.
(11) Gain calibrator.
(12) Flux calibrator.
(13) Passband calibrator.
}
\end{deluxetable*}

\begin{deluxetable*}{@{}ccccccccccc@{}}
\tabletypesize{\scriptsize}
%\tablewidth{0pt}
\tablewidth{\linewidth}
\tablecaption{NGC 6946 GMC Catalog  \label{t.catalog} }
\tablehead{ 
         \colhead{ID} &
	\colhead{$\Delta$R.A.}  &
	\colhead{$\Delta$Decl.} &		
	\colhead{$V_{\rm CO}$} &
	\colhead{$R$} &	
	\colhead{$\sigma_v$} &
	\colhead{$L_{\rm CO}$} &	
	\colhead{$M_{\rm lum}$} &
	\colhead{$M_{\rm vir}$} &
	\colhead{$T$} &
	\colhead{S/N}
	\\
	\colhead{ }  &	
	\colhead{(\arcsec)}  &
	\colhead{(\arcsec)} &
	\colhead{(\kms)} &
	\colhead{(pc)} &	
	\colhead{(\kms)} &
	\colhead{($10^4$ K \kms~pc$^2$)} &
	\colhead{($10^5~M_\sun$)} &	
	\colhead{($10^5~M_\sun$)} &
	\colhead{(K)} &
	\colhead{ } 
	\\
	\colhead{(1)} &
	\colhead{(2)} &
	\colhead{(3)} &
	\colhead{(4)} &
	\colhead{(5)} &
	\colhead{(6)} &
	\colhead{(7)} &
	\colhead{(8)} &
	\colhead{(9)} &
	\colhead{(10)} &
	\colhead{(11)} 
}
\startdata
   1   &    110.1   &  101.9  &  $-47.8$  &  30.6 $\pm$ 23.0   &   1.3 $\pm$ 1.9   &    3.5 $\pm$ 2.2    &  0.9 $\pm$ 0.6   &  0.6 $\pm$ 1.7     &  1.7	  & 5.1\\
   2   &    106.9   &  103.9  & $-45.2$	 &  42.3 $\pm$ 16.2   &   2.9 $\pm$ 2.9   &   16.3 $\pm$ 10.9  &  4.4 $\pm$ 2.9  &   3.7 $\pm$ 5.8    &  3.0	  & 8.9\\
   3   &    104.9   &  112.3  & $-45.2$	 &  79.9 $\pm$ 31.7   &   3.4 $\pm$ 1.5   &   11.2 $\pm$ 5.2    &  3.1 $\pm$ 1.4   &  9.3 $\pm$ 13.6   &  2.5  & 7.4\\
$\cdots$ & $\cdots$ & $\cdots$ & $\cdots$ & $\cdots$ & $\cdots$ & $\cdots$ & $\cdots$ & $\cdots$ & $\cdots$ & $\cdots$ \\
 390  & $-6.7$   &  $-6.9$  & 173.2  &  43.8 $\pm$ 38.9   &  3.1 $\pm$ 2.1   &   9.7 $\pm$ 18.1   &  0.7 $\pm$ 1.2   &  4.3 $\pm$ 4.7  &  3.4  &   10.1
\enddata
\tablecomments{
(1) Cloud ID.
(2) R.A. offset w.r.t the galactic center 20\hr34\mn52\fs355.
(3) Decl. offset w.r.t the galactic center \plus60\arcdeg09\arcmin14\farcs58.
(4) Central velocity of the cloud.
(5) Cloud radius.
(6) Velocity dispersion.
(7) \twelveCO~(2--1) luminosity.
(8) Luminosity-based mass from $L_{\rm CO}$.
(9) Virial mass.
(10) Peak temperature.
(11) Signal-to-noise ratio (peak temperature divided by a noise of 0.33 K). We only include clouds with S/N $>$ 5 in the catalog.\\
(The table is available in its entirety in machine-readable form.)
}
\end{deluxetable*}

\section{Methodology}  
\label{s.Methodology}
\subsection{SMA and IRAM 30 m Data}
\label{s.SMA_IRAM_obs}
We performed a 10-pointing \twelveCO~(2--1) mosaic of NGC 6946 between 2010 June and October in the compact (CMP) and extended (EXT) array configurations, with baselines ranging from 6 to 180 m. Nine pointings satisfying the Nyquist sampling were aligned along the galaxy's major axis, while the tenth field was specifically pointed at a bright clump on a spiral arm. Scans were interleaved with phase calibrators. We tuned the local oscillator frequency to 226 GHz, making \twelveCO~(2--1) in the upper sideband, and \thirteenCO~(2--1) and \CeighteenO~(2--1) in the lower sideband. The 0.8125 MHz channel width corresponds to a velocity resolution of 1.05 \kms~at 230 GHz. We also acquired complementary data with the sub-compact configuration (SUB) in 2012. These data were used to calibrate the SMA and IRAM 30 m visibilities (see Appendix A for details). Information regarding the SMA observations is shown in Table \ref{t.obslog}.

The IRAM 30 m \twelveCO~(2--1) cube of NGC 6946 was presented by \cite{L09}, and we used it to fill the central $uv$ hole to recover the total flux. The interested reader is referred to the original paper for details on the observations and data reduction. 

We reduced the SMA data in the standard way with the \texttt{MIR} package. In brief, we flagged unusable visibilities and applied calibrations (system temperature, bandpass, gain, and flux). The calibrated visibilities were then merged with the IRAM 30 m cube using the \texttt{MIRIAD}. We set the SMA SUB data as the flux standard, and used it to determine the scaling factors for visibilities. To create IRAM 30 m visibilities, we deconvolved the cube with a 16\arcsec~beam, applied SMA's 55\arcsec~primary beam, Fourier-inverted the cube, selected visibilities with baselines $<$12 m to avoid poor edge response, and finally scaled the amplitudes. 

The calibrated visibilities were Fourier-transformed and deconvolved using the CLEAN algorithm, with natural weighting to improve sensitivity. Since the composite dirty beam is very non-Gaussian, convolving the CLEAN map with the Gaussian beam given by \texttt{MIRIAD} can result in spurious total flux. As discussed by \cite{K11}, the total flux will not be conserved if the solid angle of the dirty beam differs from that of the restoring beam. To have a new restoring beam that can conserve the flux, we adopted the axis ratio and the position angle of the original restoring beam, but varied the beam size to match the solid angle of the dirty beam. We used this effective beam to restore the images, and successfully conserved the flux. The combined cube has a velocity resolution of 2.6 \kms, a noise of 0.33 K, and a beam of 1\farcs64 $\times$ 1\farcs31, corresponding to 44 pc $\times$ 35 pc at 5.5 Mpc. Figure \ref{f.mom0} shows the integrated intensity map.

\subsection{Cloud Decomposition}
\label{s.decomposition}
Cloud identification was performed with the {\tt CPROPS} package \citep{RL06}, which has been employed in many extragalactic studies (e.g., \citealt{B08,W11,G12,C14,L15,U15}). Simulations have shown that this algorithm is robust when comparing its search results to the actual 3D clouds \citep{P15a}. The reader is referred to \cite{RL06} for a thorough documentation of the algorithm and definitions of cloud properties. To minimize systematics from decomposition and to facilitate comparison with other studies, we adopted almost the same {\tt CPROPS} parameters as in \cite{C14}: {\tt threshold} = 4, {\tt edge} = 1.5, {\tt minvchan} = 1, {\tt bootstrap} = 50, {\tt sigdiscont} = 0. We identified 390 GMCs with signal-to-noise ratio $>$5 (Table \ref{t.catalog}), and we further grouped them into the nuclear GMCs (galactocentric radius $R_{\rm gal}<1$ kpc) and the disk GMCs ($R_{\rm gal}>1$ kpc) because there is a distinction in gas properties at that radius \citep{RF15}. Finally, we had 185 nuclear GMCs and 205 disk GMCs. As demonstrated in Appendix B, these clouds are well separated in the position--position--velocity space. Figure \ref{f.gmcdist} shows the location of the clouds on top of the integrated intensity map and the IRAC 3.6~\micron~image from \cite{K03}.

\subsection{CO-to-$H_2$ Conversion Factor and Mass of GMCs}
Observations have revealed discrepancies regarding the CO-to-$\rm{H}_2$ conversion factor $X_{\rm CO}$ in NGC 6946. \cite{DM12} derived an $X_{\rm CO}$ similar to the Milky Way (MW) value (4.4 $M_\sun$ pc$^{-2}$ $(\text{K km s}^{-1})^{-1}$; \citealt{S13}) by assuming that GMCs were in virial equilibrium. In contrast, analyses on CO isotopologues and dust mass surface density have revealed a conversion factor 5 to 10 times below the MW value in the nuclear region of NGC 6946 (e.g., \citealt{IB01,W02,MT04,S13}). In particular, \cite{S13} demonstrated that $X_{\rm CO}$ increases radially from 1/10 MW at the center to $\sim$MW beyond $\sim$6 kpc. As discussed by \cite{S13}, because CO is optically thick, CO emission can more easily leave a cloud if the velocity dispersion of the cloud is increased owing to external pressure. As a result, in regions with high external pressure, such as the galactic center, we would expect a lower $X_{\rm CO}$ because of an increased CO emissivity. This is perhaps the case for NGC 6946. As shown in Figure \ref{f.radialvariation}, some GMCs within 1 kpc from the nucleus have an enhanced velocity dispersion $>$10~\kms, which might in turn drive the observed low $X_{\rm CO}$. 

It is thus not appropriate to adopt a universal $X_{\rm CO}$ for NGC 6946. As a result, we used MATLAB to perform a least-squares regression to Figure 22.24 in \cite{S13} to derive $X_{\rm CO}$ for \twelveCO~(1--0) from the nucleus to 5 kpc (Figure \ref{f.xco}). The fitted $X_{\rm CO}$ increases from 0.37 to 2.23 $M_\sun$ pc$^{-2}$ $(\text{K km s}^{-1})^{-1}$, or from $\sim$1/10 to $\sim$1/2 the MW value. Then, these $X_{\rm CO}$ values were divided by $R_{21}\equiv$ \twelveCO~(2--1)/\twelveCO~(1--0) = 0.7 used by \cite{S13} before converting \twelveCO~(2--1) luminosity to cloud mass. Table \ref{t.catalog} lists GMC sizes, velocity dispersions, masses, and other parameters.

\subsection{Comparison with the Galactic and Extragalactic GMCs}
\label{s.gmcextragalactic}
It is useful to compare the NGC 6946 GMCs to those in the Milky Way and other galaxies. However, such a comparison is susceptible to systematic effects because of different observing conditions and data reduction methods involved in the studies. Therefore, in this study we only compare our results to those extragalactic observations analyzed by {\tt CPROPS} as well. These include \cite{B08}, \cite{C14}, \cite{G12}, \cite{L15}, \cite{U15}, and \cite{W11}. For the Milky Way GMCs, we used the samples presented by \cite{R16} because they adopted the same cloud property definitions as those in {\tt CPROPS} to analyze the original data in \cite{D01}.

\begin{figure*}[t]
\center
\includegraphics[angle=0,width=\textwidth]{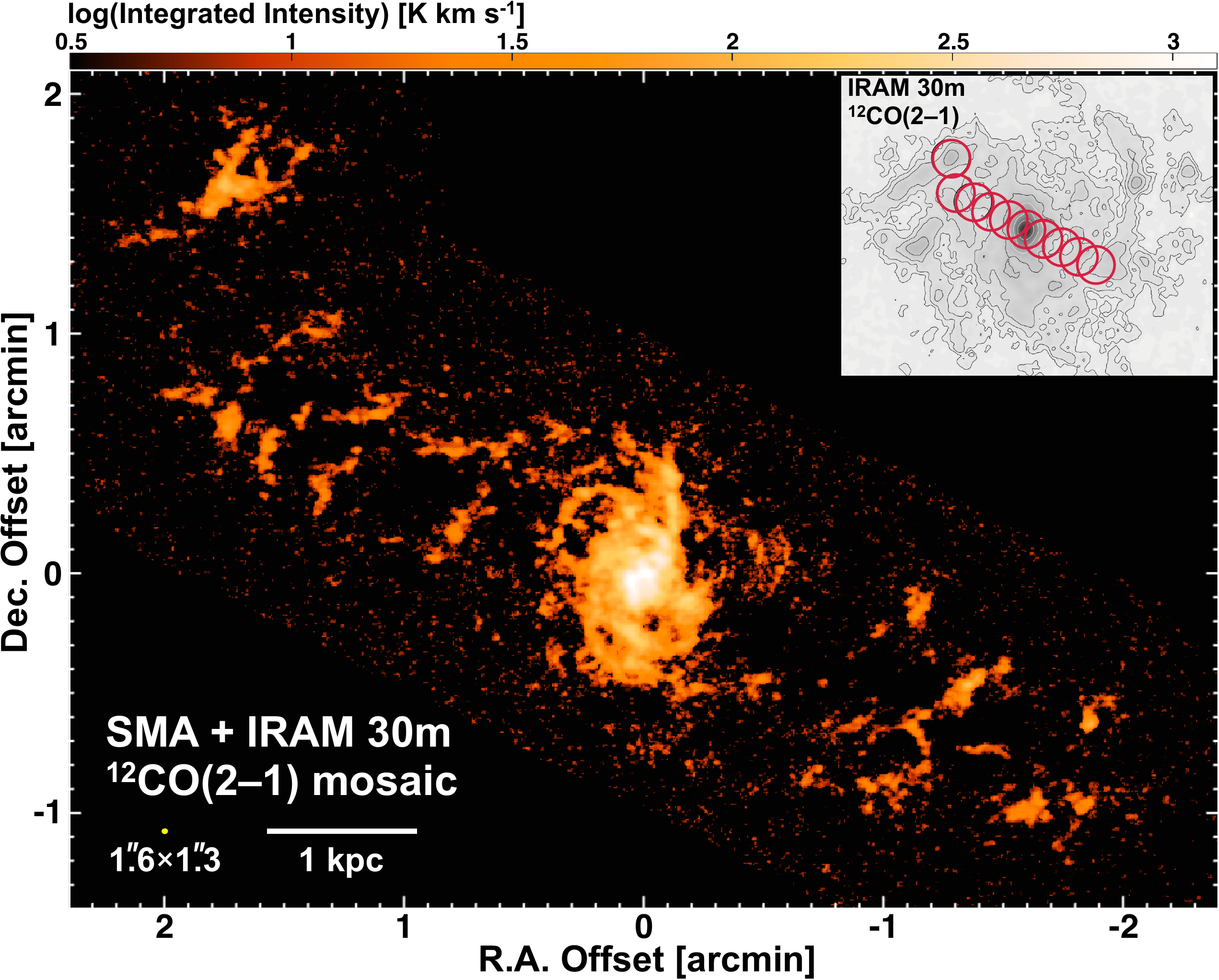}
\caption{The \twelveCO~(2--1) mosaic of NGC 6946 demonstrates a strong nuclear gas concentration and chains of GMCs in the disk. In total, 390 GMCs are identified in the data cube. The inset shows the 10 SMA pointings overlaid on the IRAM 30 m map from \cite{L09}. The $1\farcs6\times1\farcs3$ beam corresponds to 44 pc $\times$ 35 pc at a distance of 5.5 Mpc. North is up and east is left.}
\label{f.mom0}
\end{figure*}

\begin{figure*}[t]
\center
\includegraphics[angle=0,width=0.95\textwidth]{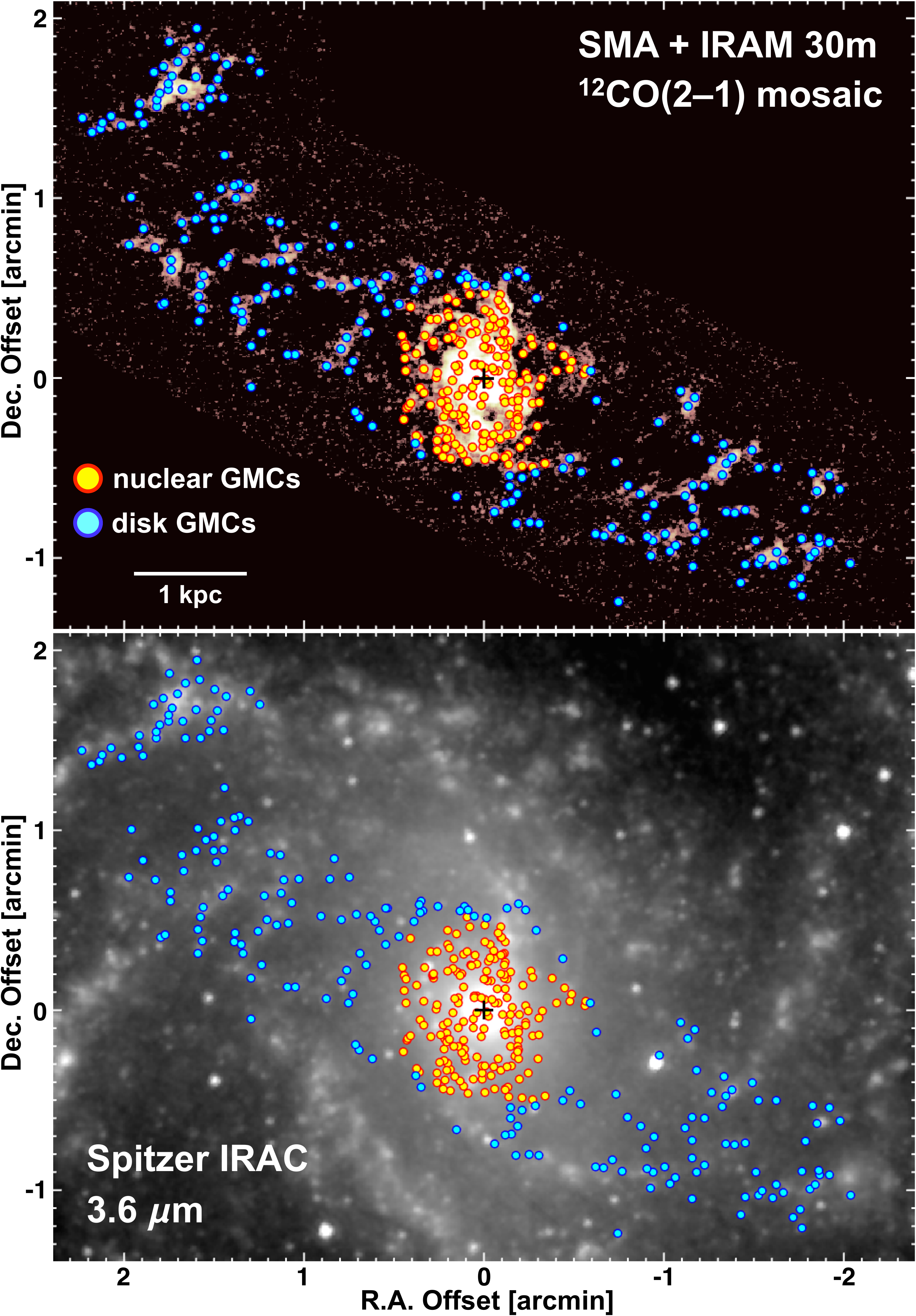}
\caption{GMCs superposed on the \twelveCO~(2--1) integrated intensity map and the Spitzer IRAC 3.6 \micron~image from \cite{K03}. The black cross marks the galactic center.}
\label{f.gmcdist}
\end{figure*}

\begin{figure}
\center
\includegraphics[angle=0,width=\columnwidth]{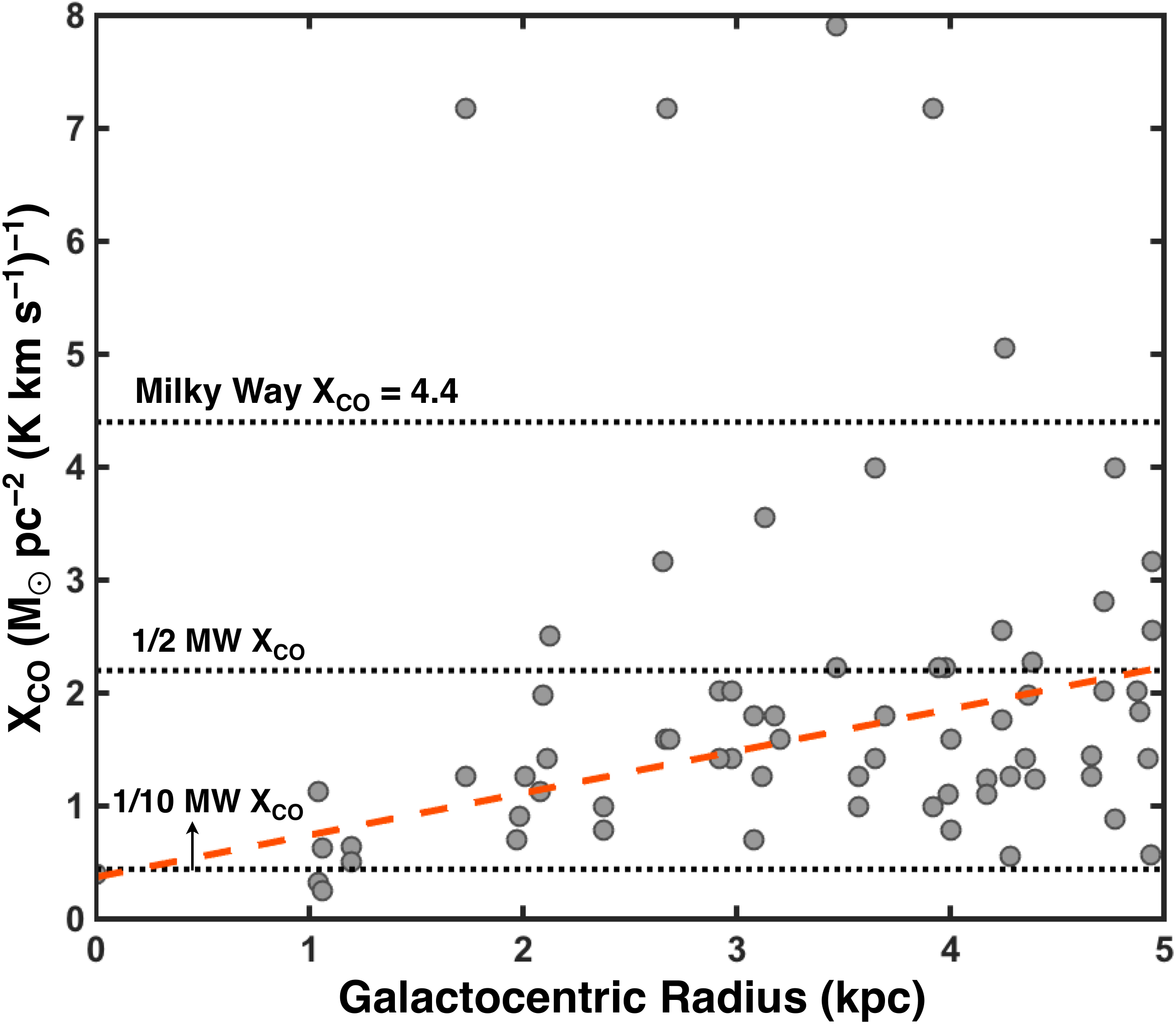}
\caption{The CO-to-$\rm{H}_2$ conversion factor $X_{\rm CO}$ increases radially in NGC 6946. Figure adapted from \cite{S13}. The orange dashed line is our linear fit to the data. The Milky Way $X_{\rm CO}$ is also plotted for comparison. To apply to our CO (2--1) data, we divided the fitted $X_{\rm CO}$ by a factor of 0.7 as adopted by \cite{S13}.}
\label{f.xco}
\end{figure}

\begin{figure}
\center
\includegraphics[angle=0,width=\columnwidth]{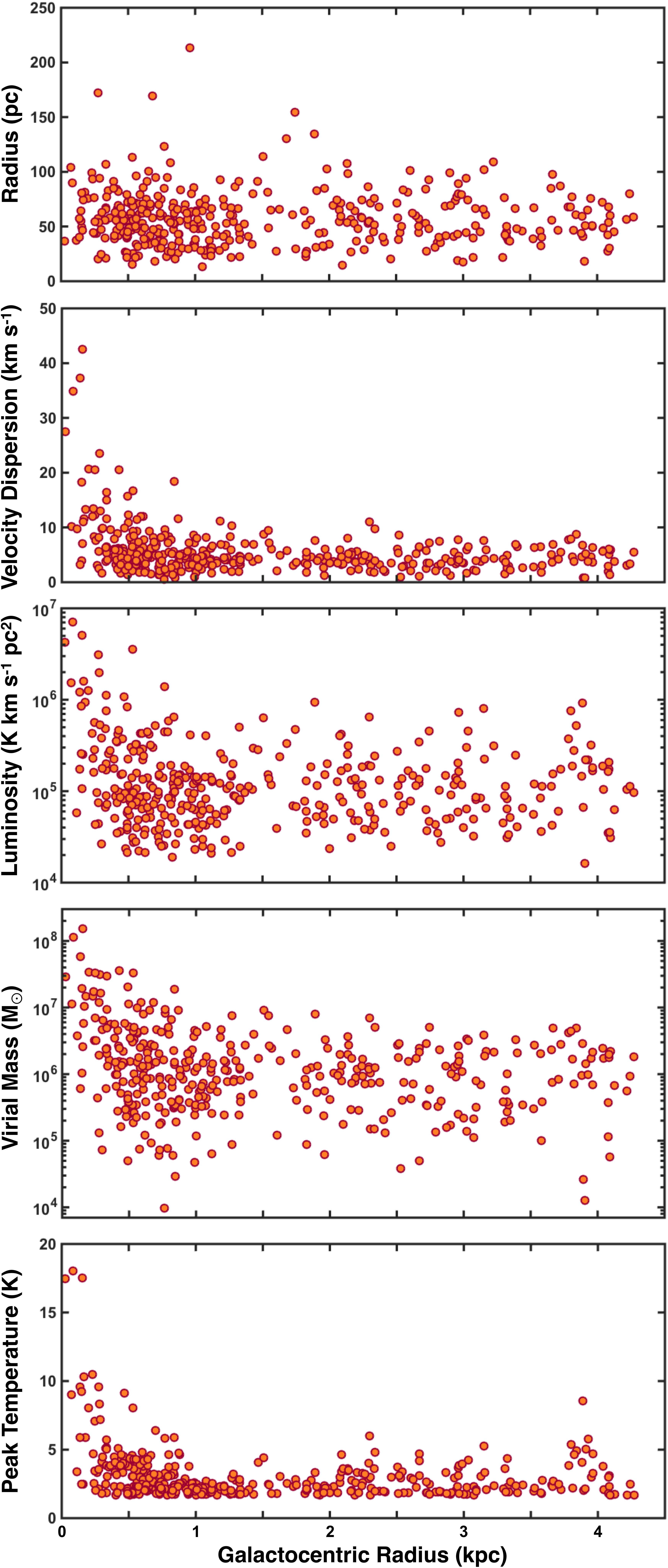}
\caption{GMC properties as a function of galactocentric radius. In NGC 6946, luminous and turbulent GMCs are predominantly located within 1 kpc from the center. Beyond 1 kpc, cloud properties are more uniform.}
\label{f.radialvariation}
\end{figure}

\begin{figure*}
\center
\includegraphics[angle=0,width=\linewidth]{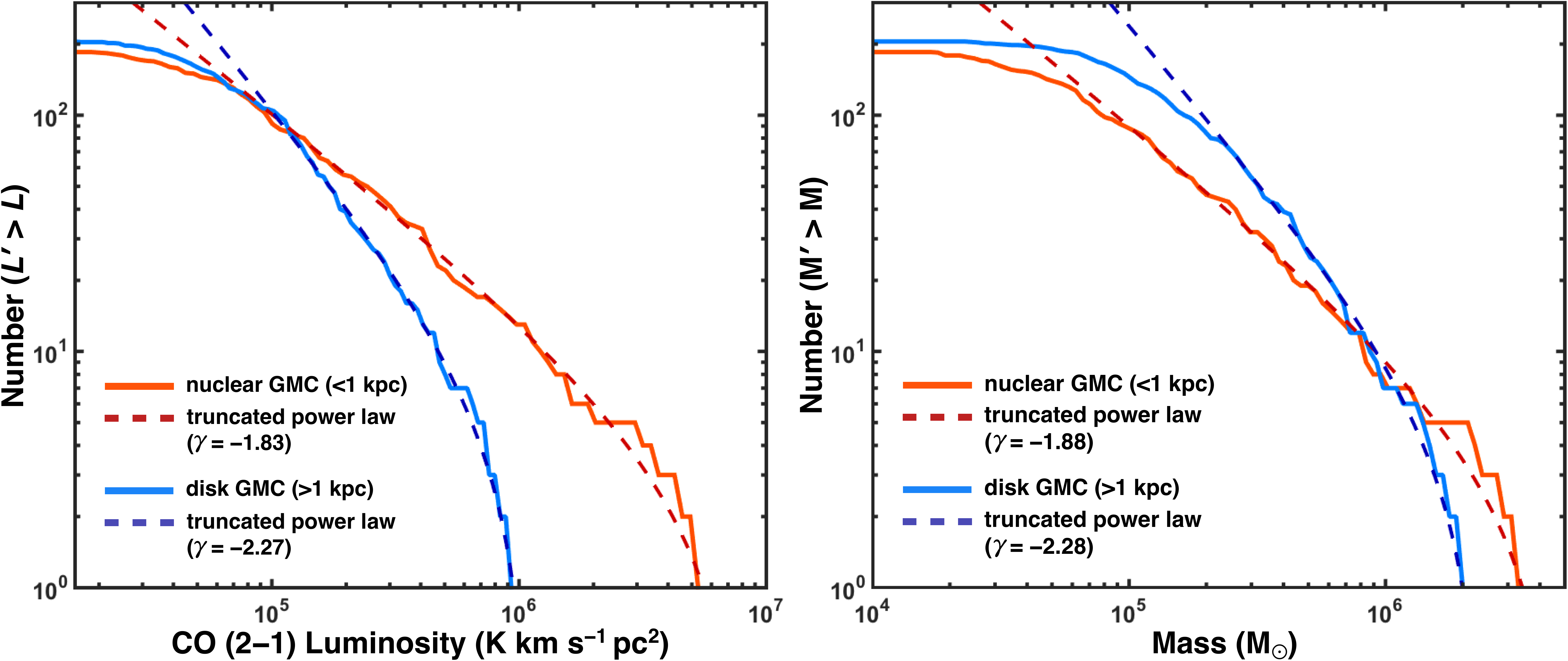}
\caption{Luminosity and mass functions for nuclear GMCs ($R_{\rm gal} <$ 1 kpc) and disk GMCs ($R_{\rm gal} >$ 1 kpc). We note that the mass function would appear identical to the luminosity function if adopting a universal $X_{\rm CO}$. Best-fit truncated power laws are shown as dashed curves. The value of the index $\gamma$ may imply different cloud formation mechanisms.}
\label{f.massfunction}
\end{figure*}

\begin{figure}
\center
\includegraphics[angle=0,width=\columnwidth]{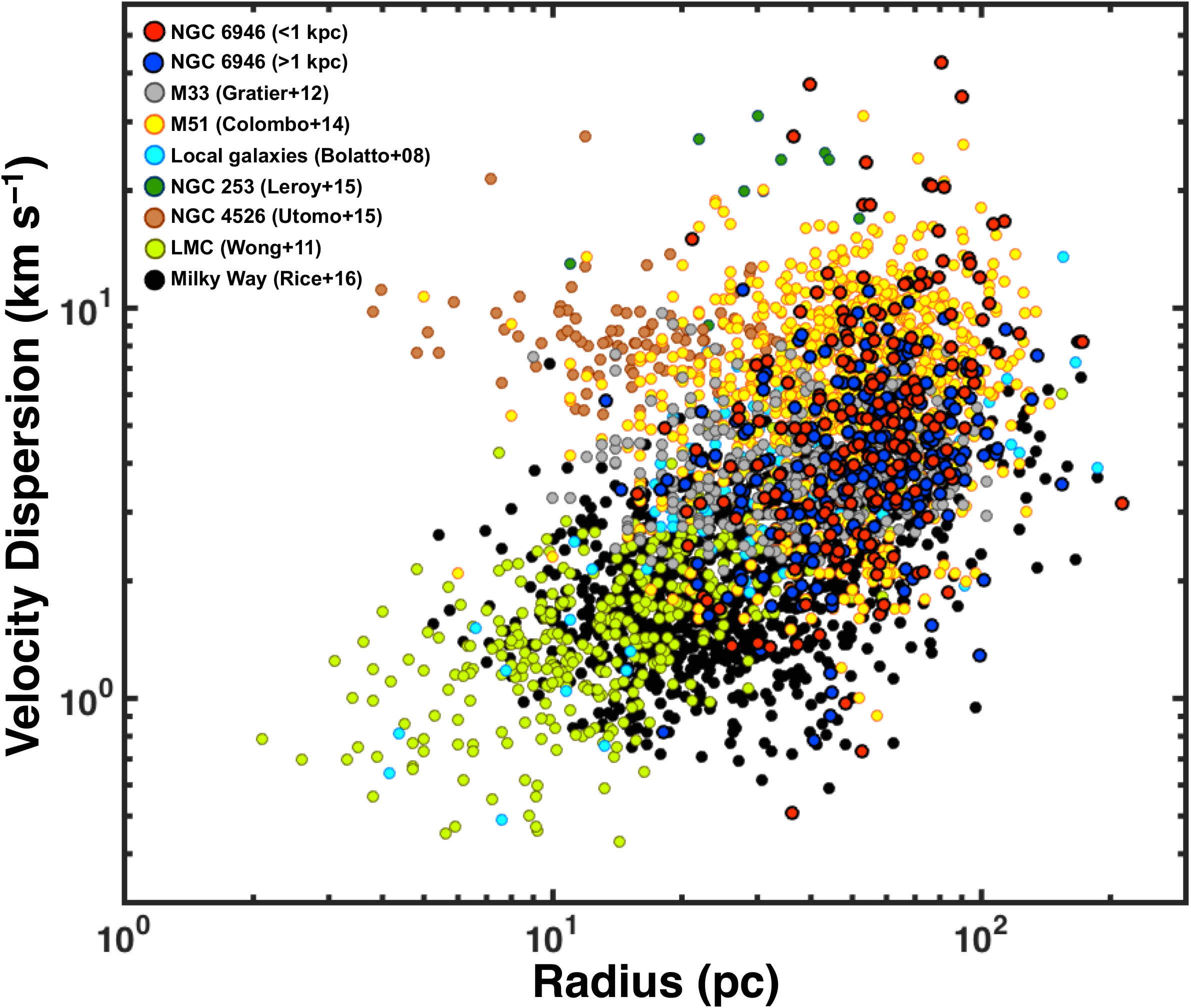}
\caption{Cloud radius vs. velocity dispersion. The NGC 6946 GMCs are very similar to those in M33, and consistent with the trend seen in the Milky Way and extragalactic clouds. Some nuclear GMCs in NGC 6946 are the most turbulent clouds ever observed, with velocity dispersions $>$10 \kms, but on average the NGC 6946 clouds are somewhat less turbulent than the M51 clouds. The mean fractional uncertainties for extragalactic clouds are 62\% for $R$ and 52\% for $\sigma_v$.}
\label{f.sizelinewidth}
\end{figure}

\begin{figure}
\center
\includegraphics[angle=0,width=\columnwidth]{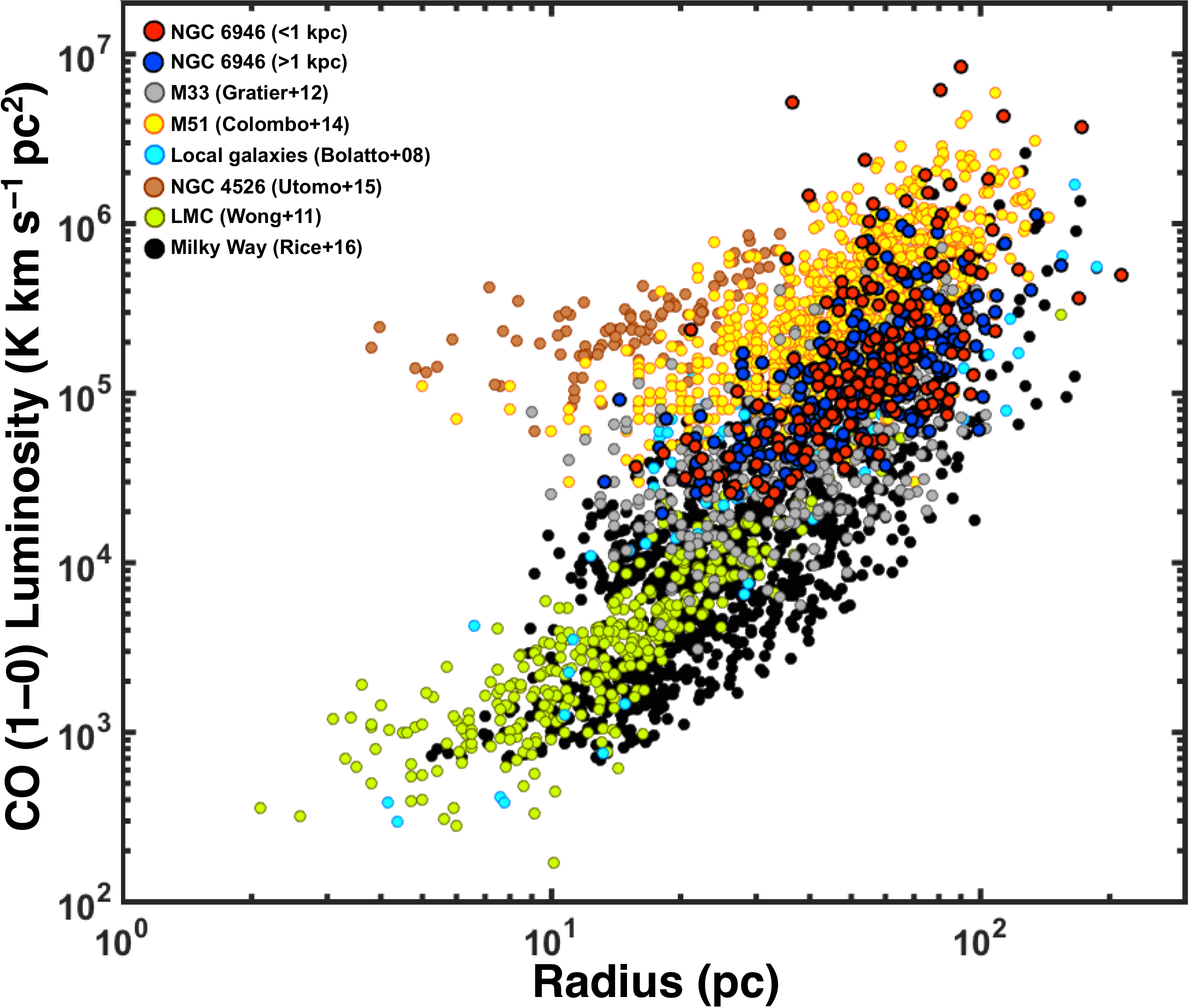}
\caption{Cloud radius vs. \twelveCO~(1--0) luminosity. We converted CO (2--1) luminosities to CO (1--0) by dividing them by $R_{21}$ ratios (see text for details). Both properties correlate well, as expected. Some nuclear GMCs are the most luminous clouds ever seen, but on average the NGC 6946 clouds are less luminous than those in M51. The mean fractional uncertainty of $L$ for extragalactic GMCs is 53\%.}
\label{f.sizeluminosity}
\end{figure}

\section{Results and Discussion}
\subsection{Gas Distribution and Radial Variation of GMC Properties}
\label{s.gasprop}
Figures \ref{f.mom0} and \ref{f.gmcdist} show the distribution of molecular clouds in our surveyed region. Molecular gas is particularly abundant in the nuclear region as the bar drives gas inward and fuels the starburst \citep{S06}. In the disk, molecular gas forms filamentary structures. While these chains of GMCs are mainly found in spiral arms (e.g., \citealt{R12}), some of them are present in the inter-arm regions.

Recently, \cite{RF15} suggested that NGC 6946 may have a ``double molecular disk'' because there is a clear transition in gas properties at 1 kpc due to the influence of the bar. They found that the inner 1 kpc is physically distinct from the rest of the disk with an exponential increase in surface density and velocity dispersion (see their Figure 3). As suggested by \cite{MT04}, molecular clouds in the nucleus of NGC 6946 are likely influenced by the nuclear bar and therefore have peculiar CO line ratios. In Figure \ref{f.radialvariation}, we show GMC properties as a function of the galactocentric radius from the center to $\sim$4.5 kpc. It is clear that massive, luminous, and turbulent clouds appear preferentially in the inner 1 kpc. Some extreme clouds can reach $>$$10^6$~K \kms~{\rm pc}$^2$ in luminosity, $>$10~\kms~in line width, and $>$$10^7$ \Msol~in virial mass. On the other hand, GMCs beyond 1 kpc tend to be more quiescent and show little radial variation in their properties, consistent with the findings in \cite{R12}. Our results provide further support to the dynamically influenced nature of the nuclear GMCs.

We note that dynamically influenced clouds have been observed in other galaxies. For example, in the barred galaxy NGC 1097, \cite{H11} found that molecular clouds on the starburst ring show an azimuthal variation in surface density and line width due to the bar-driven nuclear inflow. In the grand-design spiral M51, \cite{K09} proposed that large-scale dynamics drives coagulation on spiral arms, and fragmentation as clouds leave the arms.

\begin{deluxetable}{@{}lrr@{}}
%\tabletypesize{\scriptsize}
\tablewidth{\linewidth}
\tablecaption{GMC cumulative mass function \label{t.massfun} }
\tablehead{
	\colhead{\hspace{-15pt}Parameter}  &
	\colhead{Nuclear GMCs}  &
	\colhead{Disk GMCs}
}
\startdata  
$N_0$        &  $3.07\pm1.48$     & $4.27\pm1.93$  \\ 
$f_0/(10^6~\Msol)$       & $4.70\pm1.60$ & $2.35\pm0.46$ \\
$\gamma$   &  $-1.88\pm0.05$   &  $-2.28\pm0.09$     
\enddata
\end{deluxetable}

\subsection{GMC Luminosity and Mass Functions}
\label{s.massfunc}
Figure \ref{f.massfunction} shows that the cumulative luminosity and mass functions become steeper at the high-luminosity/mass end. To further quantify these functions, we used MATLAB to fit a truncated power law following \cite{WM97} and \cite{R05},
\begin{equation}
N(f'>f) = N_0 \Bigg[\bigg(\frac{f}{f_0}\bigg)^{\gamma+1}-1\Bigg],
\end{equation}
where $f_0$ is the maximum luminosity/mass in the distribution because $N(f'>f_0)=0$, and $N_0$ is the number of clouds more luminous/massive than $2^{1/(\gamma+1)}~f_0$, where the distribution deviates from a power law. Table \ref{t.massfun} summarizes the best-fit parameters of the GMC mass function. Compared with the luminosity function, adopting a radially increasing $X_{\rm CO}$ does not significantly change the slope of mass function. The inner GMCs are preferentially more luminous/massive, so their cumulative functions have a shallower slope ($\gamma\sim-1.9$) than that of the outer GMCs ($\gamma\sim-2.3$). 

We note that GMC mass spectra in some spiral galaxies, including the Milky Way \citep{R05,R16}, M33 \citep{R07,G12}, and M51 \citep{C14}, also have a shallower slope ($\gamma>-2$) in the inner regions, and a steeper slope ($\gamma<-2$) in the outer regions. Recently, the lenticular galaxy NGC 4526 was also found to exhibit this radial dependence of $\gamma$ \citep{U15}. The radial variation of GMC mass spectrum likely reflects intrinsic difference in the formation and evolution of molecular clouds in different regions of a galaxy (e.g., \citealt{K17}).

\subsection{GMC Formation Mechanisms}
\label{s.gmcformation}
The formation of GMCs can be largely divided into ``top-down'' and ``bottom-up'' scenarios (e.g., \citealt{MO07}). In the top-down scenario, large-scale gravitational instabilities induce the collapse of molecular gas (e.g., \citealt{E79,C81}), while in the bottom-up scenario, GMCs form by coalescence of small clouds (e.g., \citealt{K79,K09}). Simulations by \cite{D08} showed that galactic environments can determine which mechanism is dominant: self-gravity plays an important role in high surface density regions, while agglomeration seems to dominate in low-density regions where gas is stable against gravitational collapse. Furthermore, \cite{D08} demonstrated that as self-gravity becomes important, the GMC mass spectrum becomes shallower with the index $\gamma>-2$ (see their Figure 9). On the other hand, in models with no self-gravity, \cite{D08} found a steeper slope with $\gamma<-2$. These features are in line with GMCs in NGC 6946 and other disk galaxies, as summarized in the previous section. 

Therefore, both the top-down (gravitational instabilities) and bottom-up (agglomeration) scenarios may contribute to GMC formation in NGC 6946 and other galaxies. In the nuclear region of NGC 6946, large-scale bar dynamics drives a large amount of gas inward, so gravitational instabilities become effective as surface density greatly increases, thereby forming massive clouds. In the outer disk, however, surface density is not high enough to trigger instabilities, so GMCs primarily form via merging of diffuse gas or small clouds.

\subsection{Correlations of GMC Properties}
\label{s.gmccorrelations}
Here we compare the properties of the NGC 6946 GMCs to clouds in the Milky Way and other galaxies. As described in Section \ref{s.gmcextragalactic}, we only select extragalactic samples obtained with {\tt CPROPS} to minimize systematic effects. Among all the selected galaxies, our results are most directly comparable to M33 \citep{G12} and M51 \citep{C14} because of similar resolutions and sensitivities (20--50 mK noise, 2.6 \kms~channel width, $\sim$50 pc spatial resolution in \citealt{G12}; 0.4 K, 5 \kms, $\sim$40 pc in \citealt{C14}; 0.33 K, 2.6 \kms, $\sim$40 pc in our study). Moreover, we used almost the same {\tt CPROPS} parameters as \cite{C14} as listed in Section \ref{s.decomposition}. As a result, differences between statistical properties of the M33, M51, NGC 6946 clouds are more likely to be real. However, we caution that Figures \ref{f.sizelinewidth} and \ref{f.sizeluminosity} are really a hodgepodge of telescope resolutions, sensitivities, and data reduction methods, even if most of the samples were obtained with {\tt CPROPS}. In addition, galaxy inclinations may play a role in shaping the correlations \citep{P16}. Therefore, the observed slopes may not genuinely reflect the intrinsic properties of GMCs. Homogenizing these data sets is beyond the scope of this work, so we make no attempt to fit power laws to the data. We focus on qualitative comparisons.

Figure \ref{f.sizelinewidth} shows the distribution of radius and velocity dispersion for Galactic and extragalactic GMCs. The NGC 6946 GMCs on average are similar to the M33 clouds, but have lower velocity dispersions than the M51 clouds. However, some nuclear GMCs in NGC 6946 have $\sigma_v > 10$ \kms, making them the most turbulent clouds observed in extragalactic studies. 

We notice that cloud radius does not strongly correlate with velocity dispersion in individual galaxies; for example, in NGC 6946 both parameters are only mildly correlated, and in galaxies like NGC 4526 and M51 size and line width are almost independent. However, extragalactic GMCs as a whole seem to roughly follow the Milky Way scaling relation (e.g., \citealt{S87,R16}).

Figure \ref{f.sizeluminosity} compares cloud radius with CO (1--0) luminosity. We converted studies in CO (2--1) to CO (1--0) by using $R_{21}$ (the ratio between CO (2--1) and CO (1--0) fluxes) adopted in the original papers: 1 for galaxies in \cite{B08} and the Large Magellanic Cloud \citep{W11}, 0.73 for M33 \citep{G12}, 0.87 for NGC 4526 \citep{U15}, and 0.83 for NGC 6946 \citep{CT07}.

As can be seen in the figure, both parameters are strongly correlated, and the NGC 6946 GMCs closely follow the distribution of Galactic and extragalactic clouds. A handful of nuclear GMCs in NGC 6946 are the brightest clouds ever observed, with luminosities close to $10^7~\text{K km s}^{-1}~\text{pc}^{2}$. But on average, the NGC 6946 clouds are less luminous than the M51 clouds. Similar to the offset in Figure \ref{f.sizelinewidth}, the stronger spiral arms in M51 probably create more turbulent and massive clouds than other galaxies, making the M51 GMCs deviate from the rest of the samples, including the NGC 6946 clouds.

\section{Summary}
\label{s.summary}
We carried out wide-field mosaic imaging of the spiral galaxy NGC 6946 in \twelveCO~(2--1) emission with the Submillimeter Array. We merged the SMA visibilities with the IRAM 30 m data cube, and utilized {\tt CPROPS} to identify 390 GMCs in the nuclear and disk regions. The main results are as follows.

\begin{itemize}[leftmargin=0.33cm]
\item Molecular gas is particularly abundant in the inner 1 kpc, and chains of GMCs are seen in the disk. Some nuclear GMCs are among the most luminous and turbulent clouds ever observed, with $L>10^6~\text{K km s}^{-1}~\text{pc}^{2}$ and $\sigma_v>10~\kms$. Thus, GMCs in the inner 1 kpc are likely shaped by the large-scale bar dynamics. It is possible that the high-velocity dispersions of the nuclear GMCs might increase CO emissivity and therefore result in a very low CO-to-$\rm{H}_2$ conversion factor as measured by \cite{S13}. In contrast, GMCs beyond 1 kpc are fainter and more quiescent, and show little radial variation in physical properties. 

\item In NGC 6946, nuclear GMCs have a shallower mass spectrum, $dN/dM\propto M^{-1.9}$; in contrast, disk GMCs have a steeper mass spectrum, $dN/dM\propto M^{-2.3}$. Many disk galaxies, including the Milky Way, M33, M51, and NGC 4526, are also found to have a shallower slope (index $>-2$) in the inner regions, and a steeper slope (index $<-2$) in the outer regions. Compared to the simulations in \cite{D08}, we suggest that in the inner 1 kpc of NGC 6946, gravitational instabilities are more important in cloud formation and result in many massive clouds. Beyond 1 kpc, GMCs are mainly formed by agglomeration of small and diffuse clouds, leading to a steeper mass spectrum.

\item Cloud radius and velocity dispersion are not necessarily well correlated in individual galaxies; however, extragalactic GMCs as a whole seem to follow a Galactic size--line width relation. On the other hand, cloud radius and luminosity are well correlated even for individual galaxies. Compared to the M33 and M51 clouds, the NGC 6946 clouds are similar to the M33 ones, but they have lower luminosities and velocity dispersions than the M51 clouds. The strong spiral arms in M51 may help form GMCs more luminous and turbulent than other disk galaxies.

\end{itemize}

\acknowledgements
Y.-L.W. is very grateful to Jing-Hua Lin at every stage of this research project. Y.-L.W. also thanks Yue Wang for discussions. 
We thank the referee for useful comments and suggestions that significantly improved this paper. This research made use of the NASA/IPAC Extragalactic Database (NED) and NASA's Astrophysics Data System (ADS). Y.-L.W. is supported in part by the NSF AAG program and the TRIF fellowship. K.S. is supported by the Taiwanese National Science Council grants 99-2112-M-001-011-MY3 and 105-2119-M-001-036.

\section*{{\normalsize A\lowercase{ppendix} A \\ V\lowercase{isibility} A\lowercase{mplitude} C\lowercase{orrections}}}
\label{s.viscal}
\begin{figure*}[t]
\centering
\includegraphics[angle=0,width=\linewidth]{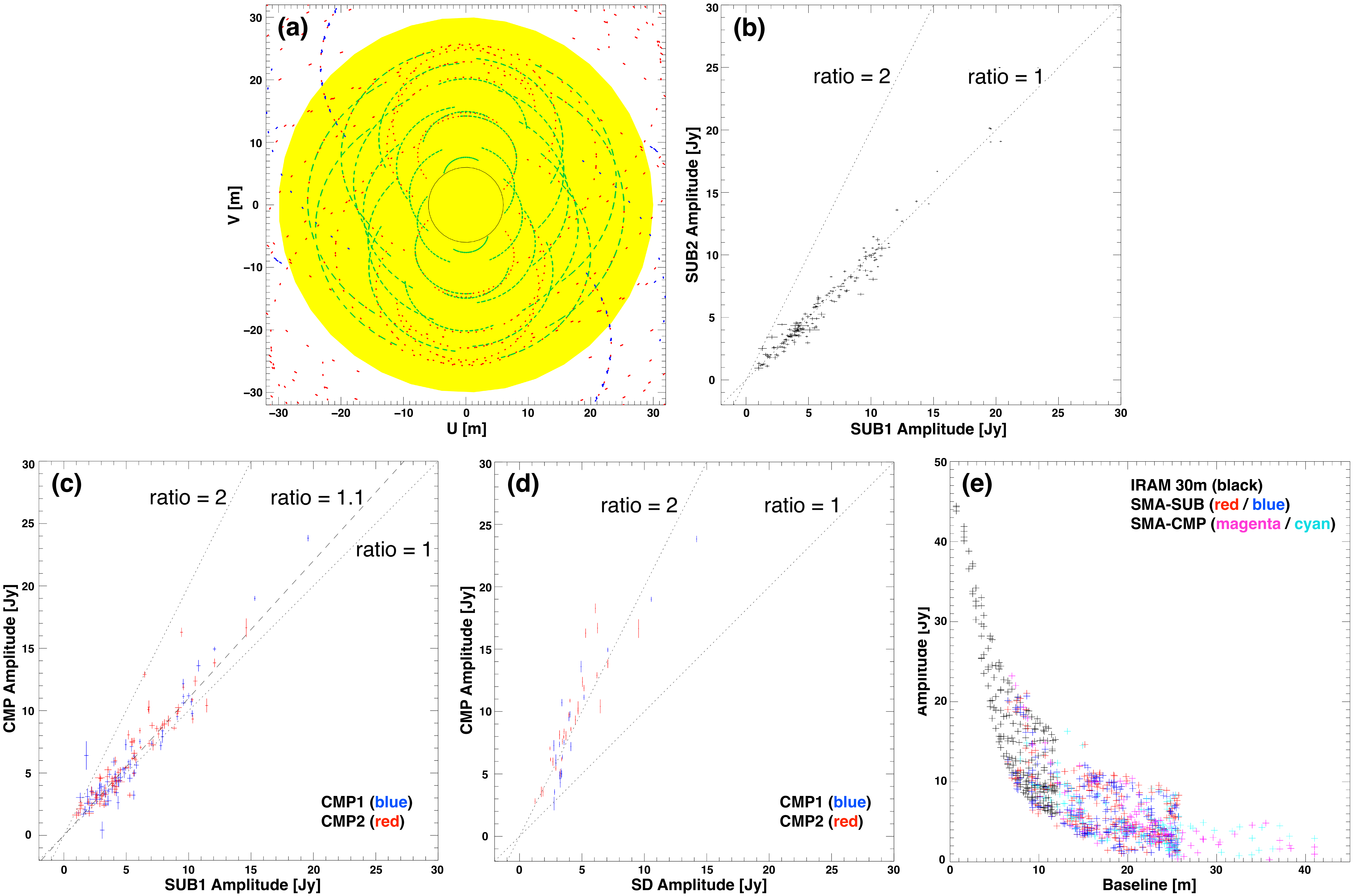}
\caption{(a) $uv$ coverage of the SMA data (red: CMP, green: SUB, blue: EXT) and the IRAM 30 m (yellow). (b) Consistency in the SUB data from two different nights suggests accurate flux calibrations. (c) Data from two CMP tracks are about 10\% brighter than that from the first SUB track. This implies that systematic error in our SMA data reduction is only at the 10\% level. (d) SMA CMP visibilities are approximately twice as bright as that of the IRAM 30 m. (e) SMA and IRAM 30 m visibilities after amplitude corrections.}
\label{f.viscal}
\end{figure*}

Here we describe the calibrations of SMA and IRAM 30 m visibilities. Panel (a) of Figure \ref{f.viscal} shows the $uv$ coverage for SMA and IRAM 30 m. The complementary SUB data ensure enough overlap between both telescopes, especially for baselines $<$10~m where the single-dish data have higher fidelity. Panel (b) shows that the SUB visibilities in both tracks are very consistent, and panel (c) shows that the CMP visibilities are, on average, $\sim$10\% brighter than the SUB ones. It appears that our SMA flux calibrations were accurate to about 10\% uncertainty; as a result, we adopted these SUB visibilities as the absolute flux standard.

In contrast, there is a significant discrepancy between the visibility amplitudes of SMA and IRAM 30 m. As panel (d) demonstrates, SMA visibilities are approximately twice as bright as those of IRAM 30 m. Inaccurate models of the single-dish beam and/or the interferometer primary beam may result in this amplitude mismatch, but the true cause remains elusive. To proceed, we determined the single-dish beam size by trial and error until the amplitude ratio was closest to a constant at short baselines. We found that the optimum size was 16\arcsec, and the ratio was $\sim$1.14 for baselines $<$12 m. We hence applied the following corrections to SMA and IRAM 30 m visibilities. (1) Divide the CMP1 data by 1.13, and CMP2 by 1.12. (2) Use the IRAM 30 m  data at short baselines ($<$12 m). (3) Adopt a larger single-dish beam of 16\arcsec, and scale the visibilities by 1.14. Panel (e) demonstrates that visibility amplitudes are well-matched after corrections.\\

\section*{{\normalsize A\lowercase{ppendix} B \\C\lowercase{loud} B\lowercase{lending}}}
\label{s.blending}
\begin{figure*}[h]
\center
\includegraphics[angle=0,width=\textwidth]{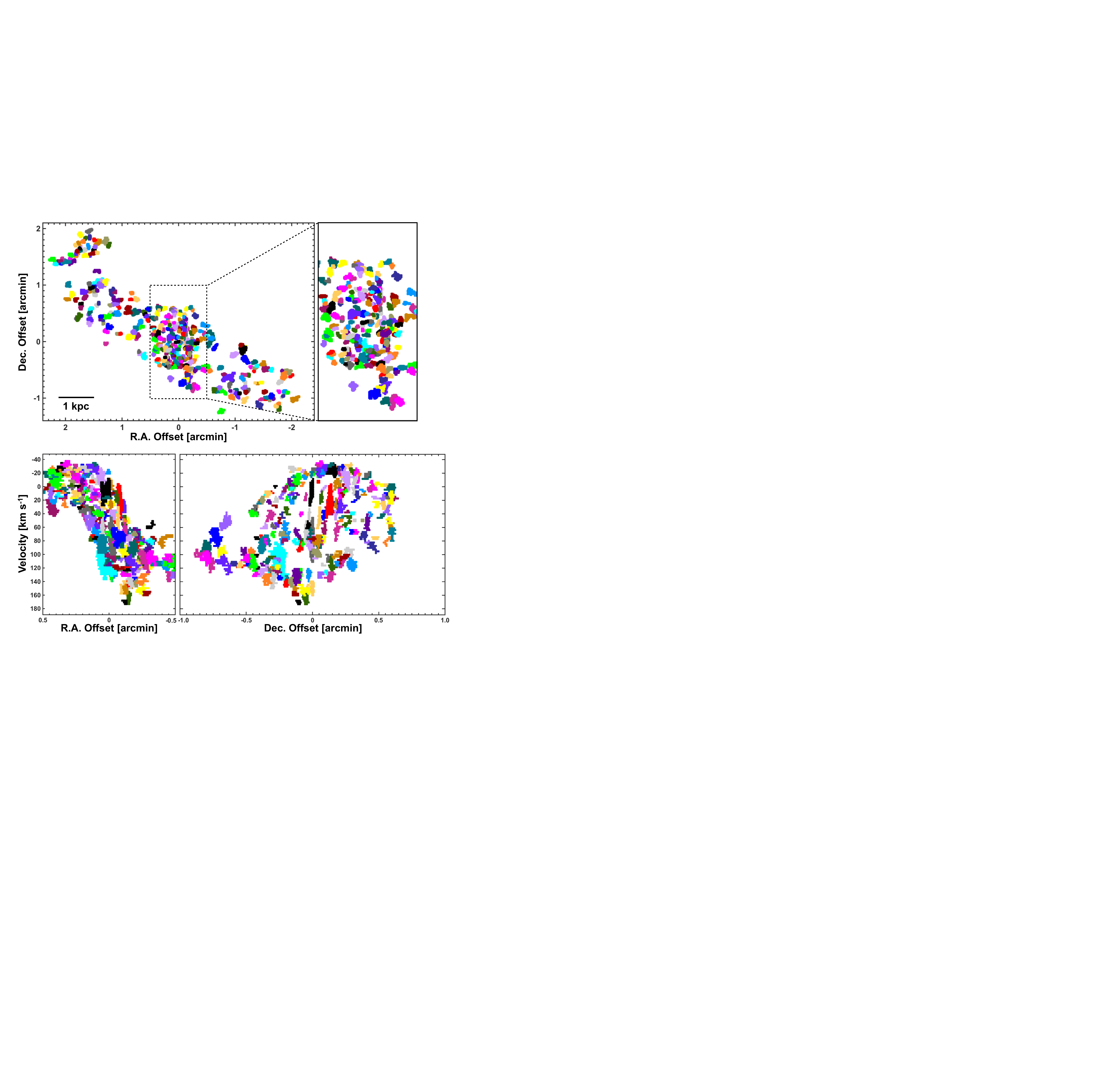}
\caption{Projections of the cloud-assignment cube on the position--position (top) and position-velocity space (bottom). Disk GMCs are generally well separated in both spaces. Some nuclear GMCs are overlapped along the line of sight, but they are still separable in the velocity space.}
\label{f.gmcblending}
\end{figure*}

Cloud blending/confusion can potentially skew our statistics; therefore, we examined the location of each GMC in the position--position--velocity space. Figure \ref{f.gmcblending} shows the ``cloud assignment cube,'' of which each color patch represents a cloud in the ppv space. GMCs in the disk region are rather separated. On the other hand, GMCs near the galactic center can look blended in the pp space, but each of them still occupies a specific location in the ppv space. We hence conclude that cloud blending/confusion is unlikely to affect our results.

\end{document}